\documentclass[letterpaper, 10 pt, conference]{ieeeconf}  % Comment this line out if you need a4paper

\IEEEoverridecommandlockouts                              % This command is only needed if 
\usepackage{graphicx}

\usepackage{subfig}

\usepackage[font={footnotesize}]{caption}

\usepackage{multicol}
\usepackage{multirow}

\usepackage[hyphens]{url}

\usepackage[acronym]{glossaries}
\glsdisablehyper

\newacronym{who}{WHO}{World Health Organization}
\newacronym{euroncap}{Euro NCAP}{European New Car Assessment Programme}
\newacronym{dms}{DMS}{Driver Monitoring System}
\newacronym{blob}{BLOB}{binary large object}
\newacronym{lac}{LAC}{looking ahead confidence}
\newacronym{svm}{SVM}{support vector machine}
\newacronym{cnn}{CNN}{convolutional neural network}
\newacronym{nir}{NIR}{near infrared}
\newacronym{ltm}{LTM}{long-term memory}
\newacronym{stm}{STM}{short-term memory}
\newacronym{drt}{DRTs}{Detection response tasks}
\newacronym{eeg}{EEG}{electroencephalogram}
\newacronym{eda}{EDA}{electrodermal activity}
\newacronym{ecg}{ECG}{electrocardiogram}

\makeatletter
\let\NAT@parse\undefined
\makeatother
\usepackage{hyperref}

\title{\LARGE \bf
Heatmap-Based Method for Estimating Drivers' Cognitive Distraction
}

\author{\authorblockN{Antonyo Musabini}
\authorblockA{\textit{InnoCoRe} \\
\textit{Valeo, Comfort and Driving Assistance}\\
Bobigny, France \\
antonyo.musabini@valeo.com}
\and
\authorblockN{Mounsif Chetitah}
\authorblockA{\textit{InnoCoRe} \\
\textit{Valeo, Comfort and Driving Assistance} \\
Bobigny, France \\
mounsif.chetitah@gmail.com}
}

\begin{document}

\maketitle
\thispagestyle{empty}
\pagestyle{empty}

\begin{abstract}

In order to increase road safety, among the visual and manual distractions, modern intelligent vehicles need also to detect cognitive distracted driving (i.e., the driver’s mind wandering). In this study, the influence of cognitive processes on the driver’s gaze behavior is explored. A novel image-based representation of the driver's eye-gaze dispersion is proposed to estimate cognitive distraction. Data are collected on open highway roads, with a tailored protocol to create cognitive distraction. The visual difference of created shapes shows that a driver explores a wider area in neutral driving compared to distracted driving. \Gls{svm}-based classifiers are trained, and 85.2\% of accuracy is achieved for a two-class problem, even with a small dataset. Thus, the proposed method has the discriminative power to recognize cognitive distraction using gaze information. Finally, this work details how this image-based representation could be useful for other cases of distracted driving detection.
\end{abstract}


\glsresetall
\section{Introduction} \label{section::Introduction}

Recent technological achievements have contributed to making vehicles greener, safer and smarter. However, despite all the efforts made regarding safety, the number of people who lose their lives due to road accidents is still rising. According to the \gls{who} road safety report from 2018 \cite{who_report}, an average of 3700 people die on the road every day, which amounts to 1.35 million victims of car crashes per year (i.e., the eighth leading cause of death of people of all ages, and the primary cause of death for children and young adults between 5 and 29 years old). The growth in the number of available vehicles on open roads is naturally a contributing factor to the rise of accident occurrences; however, the main reason is distracted driving~\cite{who_report}.

Distracted driving is described as being occupied by any activity which is unnecessary for the task of driving, such as talking or texting on the phone, eating and drinking, talking to people in the vehicle, interacting with the stereo and entertainment or navigation system---i.e., anything that takes attention away from the task of safe driving \cite{nhtsa}. Based on the \gls{who}'s source, a driver's probable distractions are clustered as follows \cite{nhtsa_faq}:

\begin{itemize}
    \item \textit{Visual distraction}: taking the eyes off the road;
    \item \textit{Manual distraction}: taking the hands of the wheel;
    \item \textit{Cognitive distraction}: taking the mind off the driving task.
\end{itemize}

Passive safety systems to combat visual and manual distraction are already widely used in commercial vehicles. These systems track the driver's eye-gaze. Once the driver looks anywhere other than the road, they are judged to be distracted \cite{US6496117B2}. The downside of this is that if the driver is looking at the road but daydreaming (a phenomenon known as the mind wandering \cite{SCHOOLER20141}), they are misjudged as attentive.

Cognitive distracted driving is a dangerous situation which vehicles should be able to detect to increase road safety. It has been highlighted as one of the issues to resolve in the \gls{euroncap} 2022 requirements (driver inattentiveness) \cite{euroncap}.

\begin{figure}[!htb]
  \centering
  \includegraphics[trim = 52mm 18mm 7mm 38mm, clip,width=\linewidth]{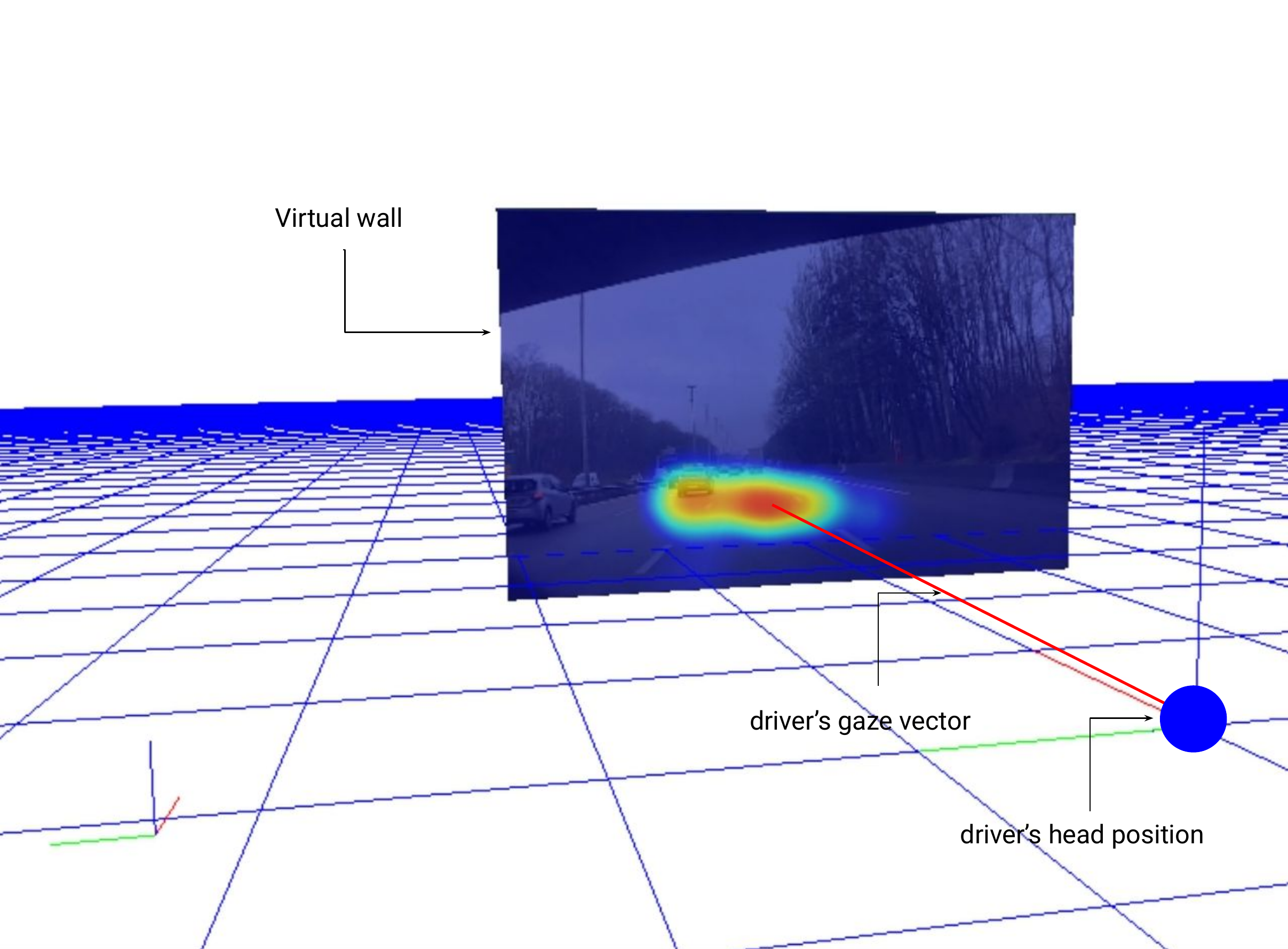}
   
  \caption{The overlay of the generated heatmap with the front camera's view, presented in a 3D illustration. The blue sphere represents the driver’s head position. The driver’s eye-gaze vector is represented as a red line protruding from the blue sphere.
  On the imaginary surface (virtual wall), the dark red spot is where the driver’s gaze activity is concentrated, and the dark blue represents the absence of any gaze activity.}
  \label{figure::3DViewerHeatmap}
\end{figure}

This work proposes to detect the cognitive load of the driver with a novel image-based representation of the driver's eye-gaze dispersion (see Figure \ref{figure::3DViewerHeatmap}), called a heatmap. Features are extracted from this representation and a \gls{svm} classifier is trained to estimate cognitive distracted driving. Additionally, the designed data collection protocol is presented. Section \ref{section::SoA} details the scientific foundation for the eye movements and the cognitive load, as well  as the state-of-the-art method; the following section, \ref{section::Methodology}, explains our experimental protocol and the data acquisition process. Then, section \ref{section::Results} presents the obtained results, and finally section \ref{section::Conclusion} presents the conclusion and further discussions.

\section{The State of The Art} \label{section::SoA}

Both biological and physiological approaches naturally influence human behavior by nature (aspects of behavior that are inherited) and nurture (aspects of behavior that are acquired). The cognitive approach deals with how people process information and how data is centered on the concept of memory by encoding, storing and retrieving information \cite{neisser2014cognitive}. Scheme, perception and working memory concepts have been proposed to reveal cognitive processes using physiological behavior.

The Multi-Store Model \cite{atkinson_shiffrin_1977} proposes that memory consists of a process including a sensory register, \gls{stm} and \gls{ltm}. \gls{stm} is developed as working memory, which is a system for temporarily storing and managing required information to carry out complex cognitive tasks such as learning, reasoning, and comprehension~\cite{Baddeley1974, 10.1007/s10648-005-3951-0}.

Cognitive load refers to the used amount of working memory resources. It is a variable which is used to assess and measure the demands on working memory and can be of the following types: intrinsic (relative complexity), extraneous (ineffective or unnecessary) and germane (effective) \cite{10.1007/978-3-030-14273-5_3}. With the increased demand on working memory placed by an abundance of novel information or by interactions of present elements, the cognitive load rises.

Existing cognitive load measurement techniques are divided into three categories; self-reports, performance measures, and physiological measures \cite{Nilsson2017}. The self-report method cannot be used as a feature by a real time vehicle application. For performance and physiological measures, numerous clues from different sources contain information about the cognitive load of the driver. For instance, a combination of vehicle data, environment data and the knowledge of the current task is used to estimate the workload placed on the vehicle driver \cite{USOO6998972B2}; the merging of the driver's eye movement, eye-gaze direction, eye-closure blinking movement, head movement, head position, head orientation, movable facial features and facial temperature image into this method has been proposed \cite{US6974414B2}. Bio-physiological signals such as driver-facing sensors and relay features such as the hands, fingers, head, eye gaze, feet, facial expression, voice tone, brain activity, heart rate, skin conductance, steering-wheel grip force, muscle activity and skin/body temperature are other signals which could be used for cognitive load estimation \cite{US10399575B2}. Other methods based on an \gls{ecg} assume that heart rhythms, controlled by the autonomic nervous system, can fluctuate with cognitive load \cite{Sahadat2013} or on the \gls{eda} \cite{8417207}. However, observing brain activity is extremely efficient to detect cognitive distraction by identifying frequency bands which are likely to capture the cognitive load and brain locations related to it~\cite{KUMAR201670}. It has been reported up to 98\% accuracy of cognitive distraction recognition, while driving is a simulator, by analyzing \gls{eeg} dynamics~\cite{7795957}. Nevertheless, this type of measurement needs numerous electrodes to be in direct contact with the head of the driver, which is not very ergonomic in a commercial vehicle (sixteen electrodes in~\cite{7795957}).

In addition, the size of the pupils increases in cases of high cognitive load, and the latter also has an impact on blinking speed \cite{vanderWel2018}. In a simulator-based experiment, the cognitive load was detected by the pupil size while the drivers were involved in spoken dialogues \cite{10.1145/1743666.1743701}. However, the blinking speed and pupil sizes are also influenced by light conditions. In a vehicle application, the cognitive distraction is also been detected by combining steering angle, vehicle speed, gaze location and head heading angle~\cite{7535416}.

Among all these available information sources, our work concentrates on a method which relies on only eye-gaze data, which is obtainable with contactlessly sensors. When the driver is distracted and experiences an increasing cognitive load, the rapid, ballistic eye movements---called saccades---of his eyes are altered, and their speed might reveal cognitive distraction. Saccades become quicker and more random with high cognitive load \cite{10.1016/j.trf.2018.01.017}.

Specific eye-related measurements such as blinks, saccades, pupils, and fixations provide a relevant and reliable assessment of cognitive load \cite{Cora2016}. An observer’s visual scanning behavior tends to narrow during periods of increased cognitive demand \cite{WANG2014227}, which is in parallel to the fact that mental tasks produce impairments of spatial gaze concentration and visual-detection \cite{Recarte2003MentalWW}. In this work, based on this knowledge, instead of detecting and analyzing all eye-related movements individually, a method which sums all the gaze activity is proposed. Thus, the driver's eye-gaze vector is projected on an imaginary distant surface. By following the temporal variation of this projection, an image-based representation is created. These shapes are expected to reveal the cognitive distraction of the driver. Similar to our study, Friedman et al. \cite{10.1145/3173574.3174226} explored another image-based representation of the movements of eye pupils (without the gaze projection on an imaginary distant surface) and achieved 86.1\% accuracy with 3D \gls{cnn}.

To the best of our knowledge, our method of gaze projection on a distant surface remains original. This method spatially represents all the summed gaze activity, i.e., where the driver looks, and can be extended with additional information, such as through the projection of the positions of other vehicles, pedestrians and road signs on the same imaginary surface (see Section \ref{subsection::FurtherDiscussions}).

\section{Methodology} \label{section::Methodology}

\subsection{Cognitive Distraction and Eye Movements} \label{subsection::CognitiveLoadandEyeMovements}

In this work, the link between short-term memory and distraction while driving is explored. Cognitive load, inattention and distraction are three different concepts. Cognitive load refers to the percentage of used resources in working memory, inattention is the state in which the driver is losing attention from the driving task to other secondary tasks, and distraction refers to the involvement of the driver in other tasks. Distraction leads to inattention from a particular task, and this causes a high cognitive load (in a driving task, this is of the germane type).

Therefore, we obtained the following assumption: during \textit{neutral driving}, the driver has sufficient cognitive resources to explore the environment and performs normal tasks related to driving, such as regularly checking the mirrors, other vehicles, road signs, etc. Among the vestibulo-ocular eye movements (fixations), saccades (rapid, ballistic movements) and smooth pursuits (slower tracking movements) should be observed \cite{Purves2001}. However, during \textit{distracted driving}, the driver has fewer cognitive resources for the driving task; thus, the gaze traces cover a smaller area. As a result, a variation of the eye movements is expected.
    
\subsection{Experimental Protocol} \label{subsection::ExperimentalProtocol}

\subsubsection{Driving Laps} \label{subsubsection::DrivingLaps}

The experimental session was composed of driving two consecutive laps on the same route (see Section \ref{subsubsection::PathandDrivingConditions}). The first round (\textit{Neutral Driving}) constituted the baseline, in which the driver performed the driving task naturally. The driver was told to relax and drive carefully. This lap was important as it allowed us to determine the baseline eye-gaze variation of the participants. The second lap (\textit{Distracted Driving}) was performed immediately after the first one: in the second lap, the driver had to perform secondary tasks (see Section \ref{subsubsection::SecondaryTasks}) designed to cognitively overload them.

\subsubsection{Path and Driving Conditions} \label{subsubsection::PathandDrivingConditions}

An important aspect of the experimental protocol was to recreate driving conditions (road, weather, traffic jams) which were as similar as possible between sessions and for both laps completed by a single participant. Therefore, a highway road near to Bobigny in France was defined as the experimental path for each participant. The speed limit on this highway was constant (90 km/h), and it took 22 minutes to complete a single lap. Driving was performed during the day-time between 10am and 5pm in order to minimize the variation in weather and traffic conditions.

\subsubsection{The Expert} \label{subsubsection::Expert}

The expert was in charge of the experiment protocol, launching the secondary tasks, annotating events and guiding the driver on the driving path. He was also in charge of momentarily pausing the secondary tasks whenever the road situation became dangerous (i.e., when another vehicle overtook the test vehicle). This expert is called the \textit{accompanist} in the following sections.

\subsubsection{User Group} \label{subsubsection::UserGroup}

Five drivers participated in the data collection protocol. All of them were volunteers working in the automotive industry; however, they were not aware of the purpose of the driving session. All the participants were male, with an average age of 29.4 years.

\subsubsection{Secondary Tasks} \label{subsubsection::SecondaryTasks}

The aim of the secondary tasks was to increase the mental workload of the driver.
In the literature, distinct secondary tasks have been cited such as foot tapping (secondary task) while learning (primary task) and measuring the rhythmic precision \cite{doi:10.1002/acp.3100} or measuring the \gls{drt} while driving \cite{Nilsson2017}. In a simulator-based experiment, drivers had to accomplish visual, manual, auditory, verbal and haptic secondary tasks. Results of the eye-glance analysis showed that the visual \gls{drt} were more efficient than the other ones \cite{10.17077/drivingassessment.1563}. A vehicle oriented study used visuospatial secondary tasks (the participants should visualize
the location of this time’s hour and minute hands on the face of
an imaginary analog clock)~\cite{7535416}. However, in our study, in order to keep the eye-gaze patterns as neutral as possible, visual and visuospatial secondary tasks were discarded. Immersive and fun secondary tasks have been designed in order to attempt to reach a more natural experimental procedure. The following four games were designed, all for the n-back task strategy. The n-back tasks are cognitively distracting tasks in which the participants have to recall successive instructions. Recalling these successive instructions increases their mental workload~\cite{10.3389/fpsyg.2018.02208}. Each game was designed to last four minutes with one minute of pause between them.

\begin{itemize}
    \item \textit{{Neither Yes nor No}}: This game was based on avoiding the words "yes", "no" and their alternatives such as "yeah", or "oui". The accompanist asked successive questions to force the participants to pronounce these words.
    \item \textit{{In My Trunk There Is}}: The game consisted of citing "In my trunk there is" followed by an item’s name. The participant and the accompanist, turn by turn, had to recall all the past objects and add a new one to the list.
    \item \textit{{Guess Who?}}: The participant thought about a real or imaginary character and the \textit{accompanist} tried to determine the identity of the character by asking questions from a mobile application. The participant had to answer the questions correctly.
    \item \textit{{The 21}}: The accompanist started to count and stated 1, 2 or 3 digits in  numerical order (e.g., 2 digits: 1, 2). The driver followed the numerical order and stated it, and added a different number of digits than the accompanist (e.g., 3 digits: 3, 4, 5). The game continued in this manner; however, it was forbidden to say the number "21". When the counter arrived to "21", instead of saying "21", a new rule had to be added to the game (e.g., do not say multiples of 4) and the counter was reset to zero.
\end{itemize}

\subsection{Data Acquisition} \label{subsubsection::DataAcquisition}

\begin{figure}[!htb]
  \centering
  \includegraphics[trim = 0mm 0mm 20mm 0mm, clip,width=\linewidth]{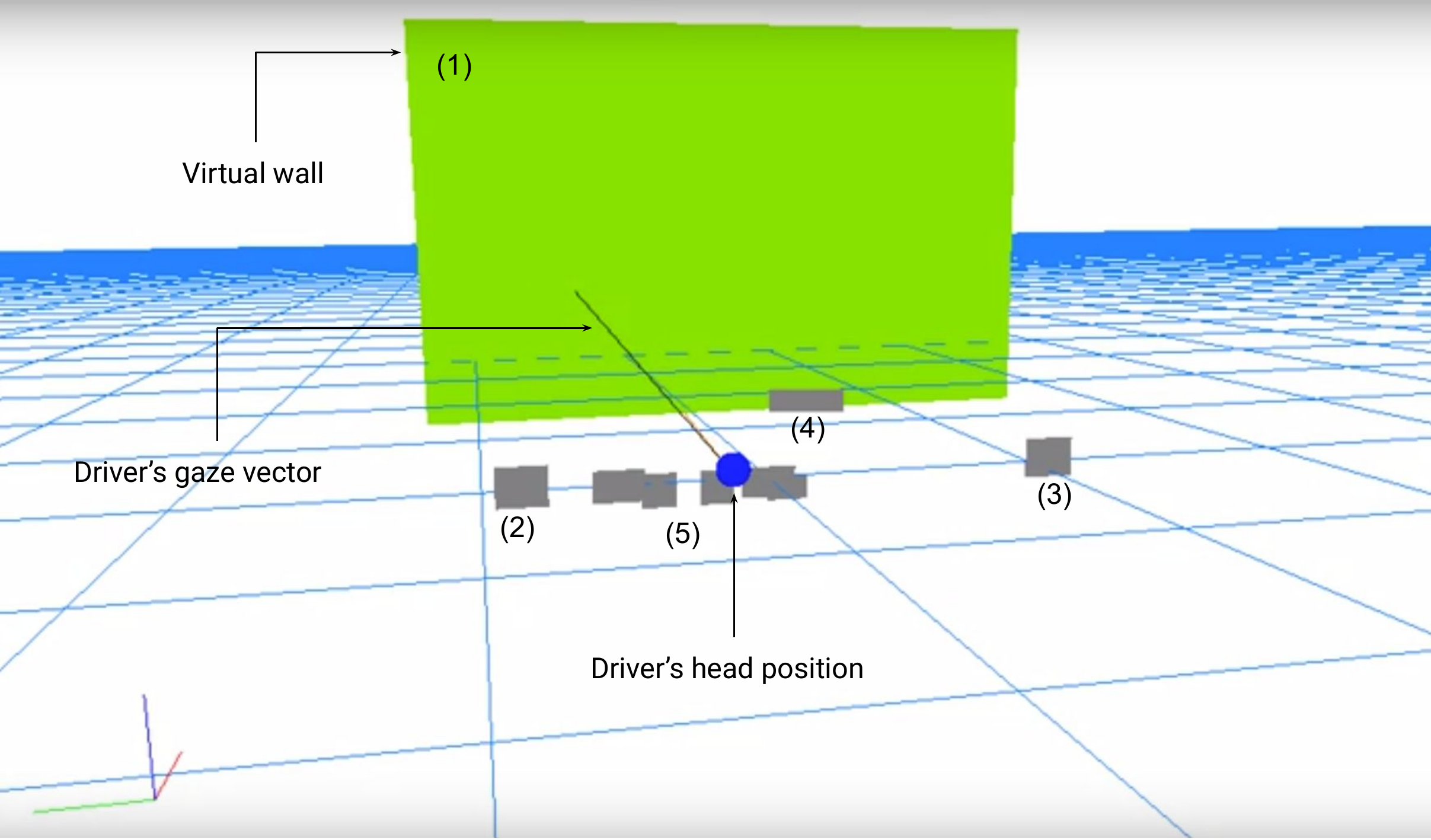}
  \caption{The 3D illustration of vehicle elements. The blue sphere represents the driver’s head position. The driver’s eye-gaze vector is represented as a black line protruding from the blue sphere. The gray rectangles represent the vehicle's interior parts: left mirror (2), right mirror (3), central mirror (4) and the instrument cluster of the speed indicator, rpm indicator, central stack and navigation screen (5). The imaginary plane surface (noted virtual wall on the figure) is illustrated as (1).}
  \label{figure::3DViewerGdi}
\end{figure}

The position of the vehicle's interior parts, such as the mirrors and the instrument cluster, were measured and illustrated in a 3D world representation (see objects 2, 3, 4 and 5 in Figure \ref{figure::3DViewerGdi}).

While driving, the driver was monitored with a \gls{nir} camera, placed in front of the instrument cluster. This sensor, part of the Valeo \gls{dms}~\cite{Valeo_dms}, extracted the head position and eye position and their direction. These data were also imported to the 3D world representation (see Figure \ref{figure::3DViewerGdi}). Thus, it was possible to detect if the driver was looking towards one of the objects present in the scene.

In addition, an imaginary plane surface was placed in front of the vehicle as if it were one of the vehicle's interior parts (object 1 in Figure \ref{figure::3DViewerGdi}). The eye-gaze vector was projected on this surface, and their intersection point was tracked for a given time window. By following the variations of the intersection point over this surface, image-based representations were generated (see Section \ref{subsubsection::HeatmapGeneration}). This representation, called a heatmap, was used to detect the cognitive load of the driver.

The vehicle was also equipped with a frontal RGB camera providing an image with a 1280 x 800 pixel resolution. The position and the dimensions of the imaginary surface were set to maximize the junction of this surface with the RGB camera's field of view and the area in which  gaze detection was available. In our  vehicle's configuration, these conditions were met when the virtual wall was placed  4 meters in front of the vehicle (point zero was selected the navigation screen of the car). Then, the vehicle was physically placed in front of a real wall, at the computed distance, and the camera's field of view was measured in meters (4.15 m x 2.59 m). In conclusion, the first step of the data acquisition process was to detect the location of the projection of the eye-gaze on the 3D imaginary surface (which was 4.15 m x 2.59 m) and convert it to pixels (i.e., 1280 x 800). The generated heatmaps were down-sampled to 640 x 400 to increase computational speed.

The RGB camera was located at the center of the vehicle, whereas the driver was sitting on the front left seat. Thus, the driver's gaze activity seemed to be concentrated on the left side of the image on the overlays and heatmaps.

\subsection{Heatmap Generation} \label{subsubsection::HeatmapGeneration}

The heatmap is a data visualization technique used in different studies and solutions. Heatmaps are often used to highlight areas of interest; therefore, we can explore several situations which arise from it. The heatmap (visible in Figure \ref{figure::3DViewerHeatmap} was used for both visualization and feature extraction after performing the following steps:

\begin{figure}[!htb]
     \centering
     \subfloat[Buffered intersection points.]{\includegraphics[width=3cm]{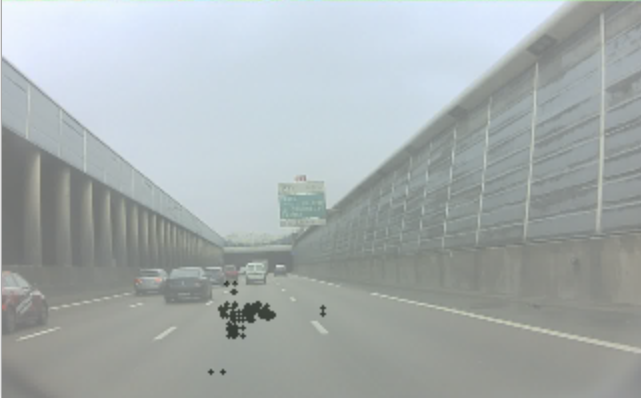}\label{figure::1pixel}}
     \qquad
     \subfloat[Field of View adaptation]{\includegraphics[width=3cm]{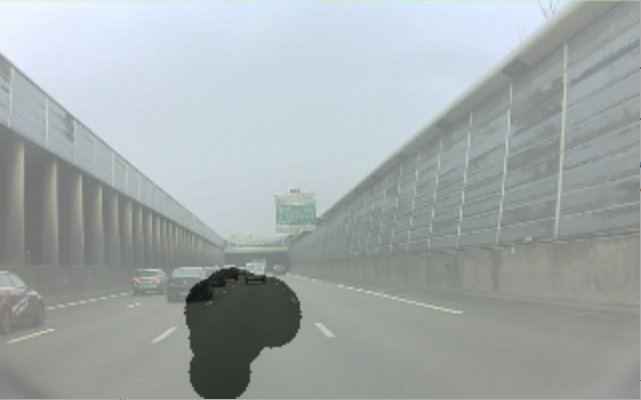}\label{figure::15pixel}}
     \qquad
     \subfloat[Opacity variation]{\includegraphics[width=3cm]{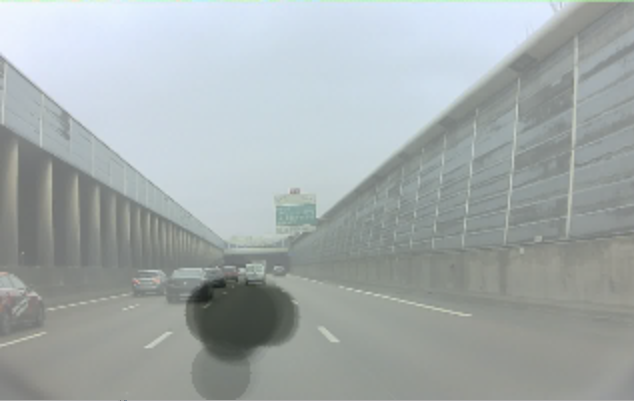}\label{figure::opacity}}
     \qquad
     \subfloat[Blurred mask]{\includegraphics[width=3cm]{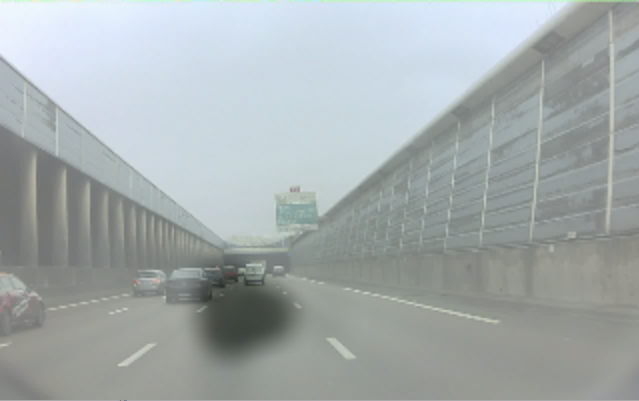}\label{figure::gaussian}}
     \caption{Successive steps for creating a heatmap mask.}
     \label{figure::heatmapSteps}
\end{figure}

\begin{itemize}
    \item {Point acquisition}: The timestamped raw intersection points for x and y between the eye-gaze vector and the imaginary surface were the heatmap generator's input. These data were acquired every 50 ms, if the driver was looking through this imaginary plane (if the driver was not looking through the plane---i.e., checking his phone---see Section \ref{features::Otherfeatures}).
    \item {Buffering---window size}: A single intersection point was not sufficiently meaningful for this specific problem. Therefore, the points were buffered as sliding windows (see Figure \ref{figure::1pixel}). Section \ref{subsubsection::Scores} compares 12 window sizes from 5 to 60 seconds.
    \item {Field of view}: With the aim of covering the field of view of the driver, a circle of 15 pixels was placed, centered on the intersection points (Figure \ref{figure::15pixel}). The choice of the circle diameter that represents the gaze fixation was mainly influenced by the pixel dimensions of our heatmaps (640 x 400).
    \item {Opacity}: After the normalization of the field of view circles, the obtained mask was used to vary the opacity of intersections (see Figure \ref{figure::opacity}).
    \item {Blurring}: Finally, a Gaussian filter was applied to reduce the noise due to the gaze activity and to concentrate on the most explored area (see Figure \ref{figure::gaussian}).
\end{itemize}

\subsection{Feature Extraction} \label{subsubsection::FeatureExtraction}

Feature engineering was applied on the generated heatmaps in order to reduce the data dimension. From each heatmap, the following feature sets, based on their pixel intensities and shape, were extracted:

\subsubsection{Appearance Features} \label{subsubsection::AppearanceFeatures}

The pixel intensity variation of a heatmap contains information on the area checked by the driver. The histogram is an efficient tool to visualize the data distributions.

\begin{figure}[!htb]
     \centering
     \subfloat[Pixel intensities]{\includegraphics[width=4.3cm]{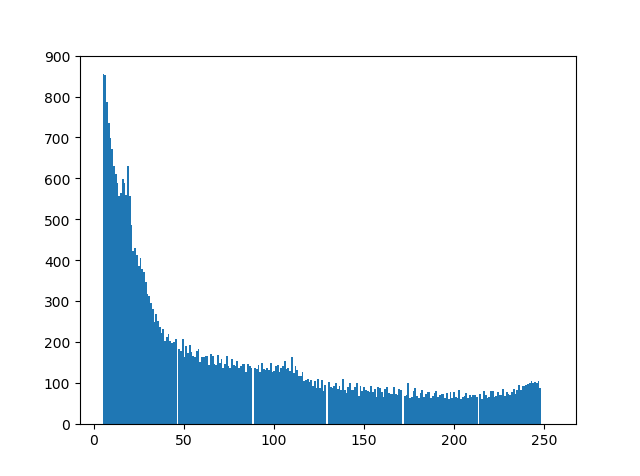}\label{figure::histogram_v1}}
     %\qquad
     \subfloat[Histogram Bins]{\includegraphics[width=4.3cm]{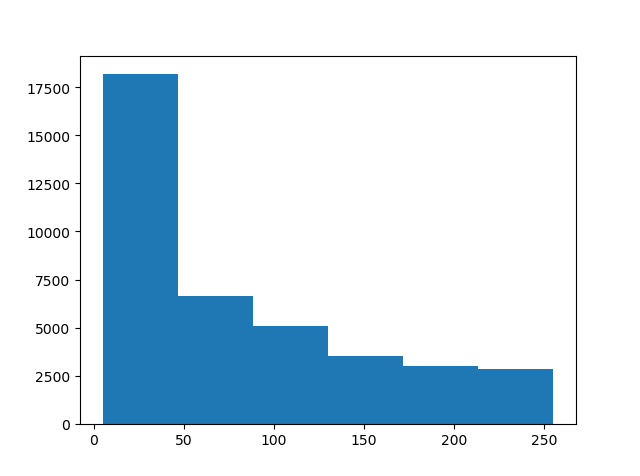}\label{figure::histogram_bins}}
     \caption{Computed histogram of the heatmap presented in Figure \ref{figure::3DViewerHeatmap}. (a) Pixel intensities are represented on the abscissa, and the number of pixels on the ordinate. (b) The values from $a$ are distributed into 6 bins.}
     \label{subplot::hist}
\end{figure}

During \textit{distracted} driving, it is expected that we see a higher concentration on higher intensities than during \textit{natural} driving, as the driver should cover a wider area, it is expected that the histogram should exhibit a shift towards low-intensity bins. Hence, a six-bin-histogram of the pixel number in terms of pixel intensity is generated per heatmap (see Figure \ref{subplot::hist}).

\subsubsection{Geometric Features} \label{subsubsection::GeometricFeatures}

\begin{figure}[!htb]
     \centering
     \subfloat[Driver’s gaze activity heatmap.]{\includegraphics[width=3cm]{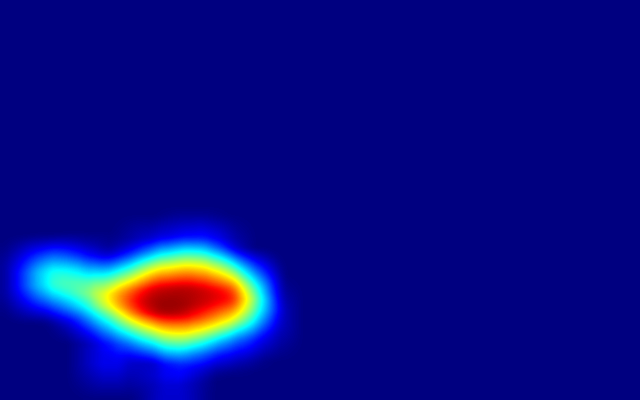}\label{figure::colorMap}}
     \qquad
     \subfloat[Thresholded heatmap]{\includegraphics[width=3cm]{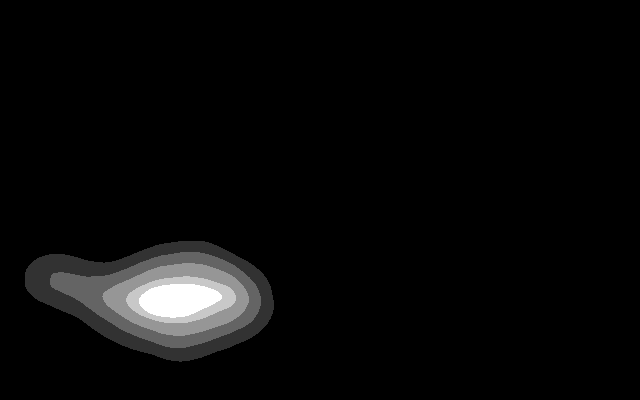}\label{figure::threshold6}}
     \qquad
     \subfloat[Zone 1]{\includegraphics[width=3cm]{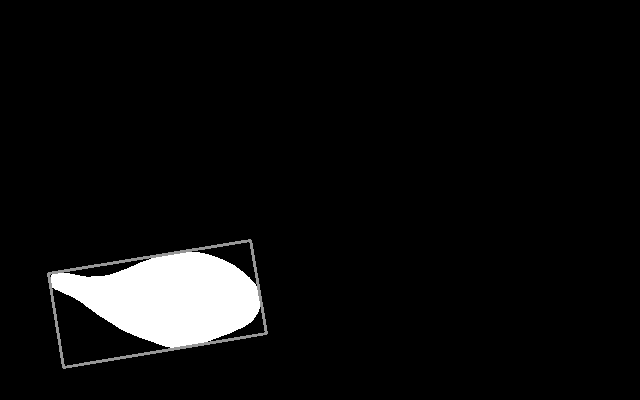}\label{figure::contourZone1}}
     \qquad
     \subfloat[Zone 2]{\includegraphics[width=3cm]{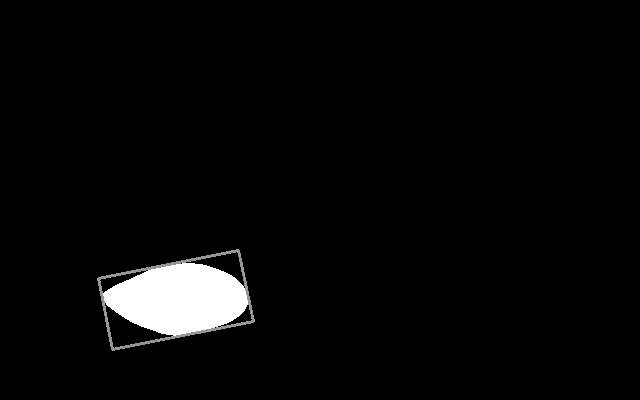}\label{figure::contourZone2}}
     \qquad
     \subfloat[Zone 3]{\includegraphics[width=3cm]{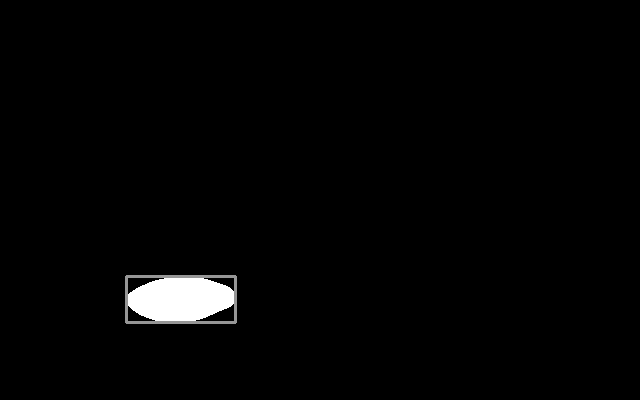}\label{figure::contourZone3}}
     \qquad
     \subfloat[Zone 4]{\includegraphics[width=3cm]{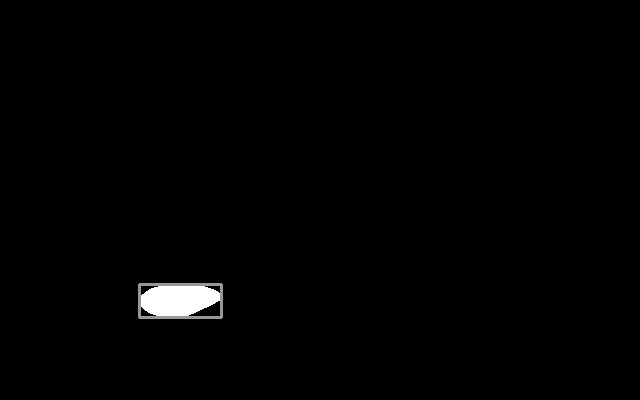}\label{figure::contourZone4}}
     \caption{Gaze activity \gls{blob}s. (a) Driver’s gaze activity heatmap. The red area represents the most fixated area and the blue region is the less fixated area. (b) Thresholded heatmap, converted to a grayscale image with distinguished contours of focus. The following figures are extracted contours from the thresholded heatmap. Each contour is defined by the pixel intensities. A binary threshold is performed for each zone.}
     \label{subplot::BlobZones}
\end{figure}

Beyond the raw pixel intensities, during distracted driving, the dispersion of the gaze activity is expected to vary differently on the abscissa and ordinate axes. Thus, their geometric form also has to be considered. The generated heatmap is divided into contours according to the differences in pixel intensities: \gls{blob}s (see Figure \ref{subplot::BlobZones}).

In order to understand the information about the driver's gaze dispersion across the imaginary plane, the following features are extracted as statistical measures from all \gls{blob}s:

\begin{itemize}
    \item Standard deviation on $x$ and $y$;
    \item Coordinates of the centroid;
    \item Boundaries of each zone (min. and max. of $x$ and $y$);
    \item First quartile, median and third quartile on $x$ and $y$;
    \item Area of the contour;
    \item Perimeter of the contour.
\end{itemize}

\subsubsection{Looking Ahead Confidence} \label{features::Otherfeatures}
If the driver does not always look through the imaginary plane during the heatmap generation time window (i.e., they are engaging in activities such as checking their phone) or if the camera is not able to detect the driver's gaze (i.e., the driver might cover the camera with his arm while manipulating the steering wheel), the observation will contain less relevant data. Therefore, the information regarding how much time the driver spent looking ahead is another feature which determines the quality of that heatmap, called \gls{lac}.

Finally, all the extracted features are standardized by removing the mean and scaling to unit variance per heatmap.

\subsection{Classifier Training} \label{subsubsection::Classifier}

The \gls{svm} supervised binary classification algorithm is trained with the extracted features (scikit-learn implementation on Python computer language~\cite{scikit-learn} with radial basis function kernel). Data collected during neutral driving have been annotated as \textit{neutral} and data collected during the secondary tasks have been annotated as \textit{distracted}.

The classification is validated through a stratified k-fold cross-validation technique, with 10 iterations ($k=10$). The leave-one-driver out technique is used to ensure the test data are always different from the training data. Stratification seeks to ensure that each fold is representative of all strata of the data, which aims to ensure that each class is equally represented across each test fold and consists of splitting the data set into samples.

\section{Results} \label{section::Results}

\subsection{Shape Visualization} \label{subsubsection::HeatmapVisualisation}

\begin{figure}[!htb]
     \centering
     \subfloat[5 seconds, distracted]{\includegraphics[width=3cm]{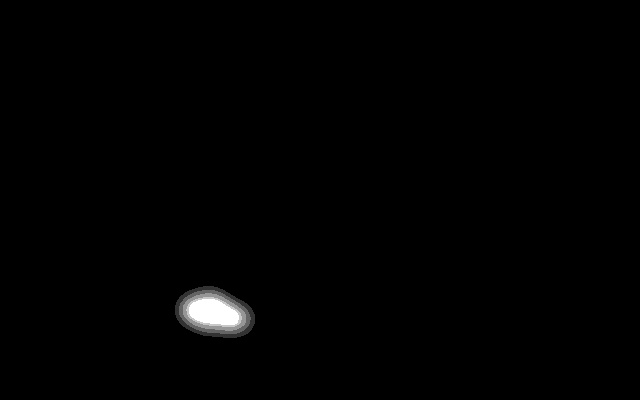}\label{figure::5D}}
     \qquad
     \subfloat[5 seconds, neutral]{\includegraphics[width=3cm]{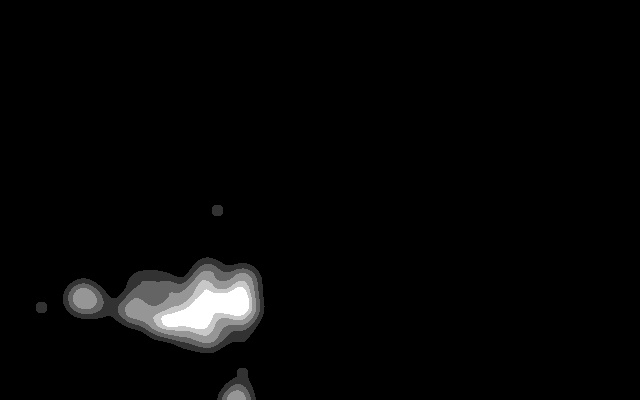}\label{figure::5N}}
     \qquad
     \subfloat[15 seconds, distracted]{\includegraphics[width=3cm]{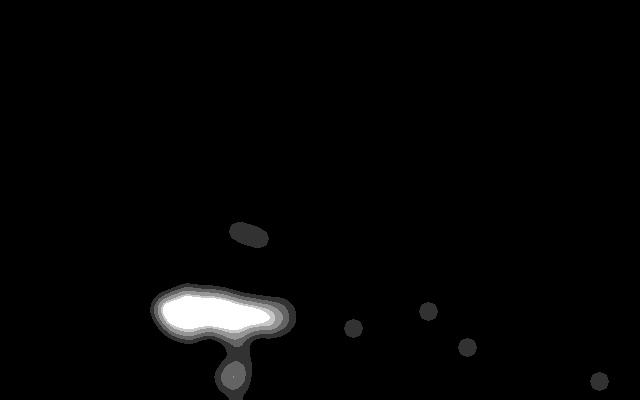}\label{figure::15D}}
     \qquad
     \subfloat[15 seconds, neutral]{\includegraphics[width=3cm]{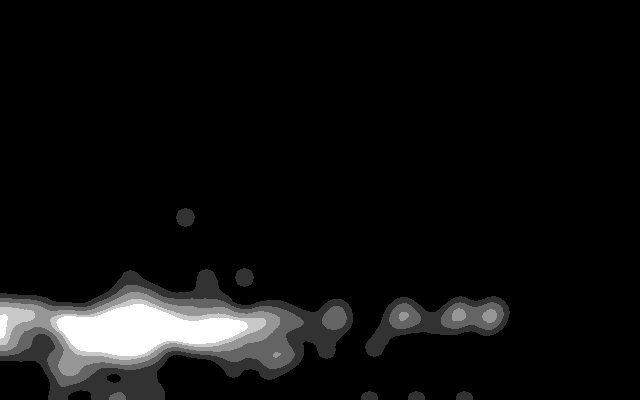}\label{figure::15N}}
     \qquad
     \subfloat[30 seconds, distracted]{\includegraphics[width=3cm]{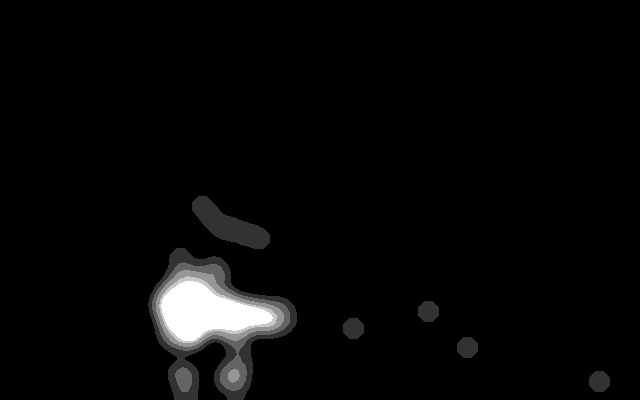}\label{figure::30D}}
     \qquad
     \subfloat[30 seconds, neutral]{\includegraphics[width=3cm]{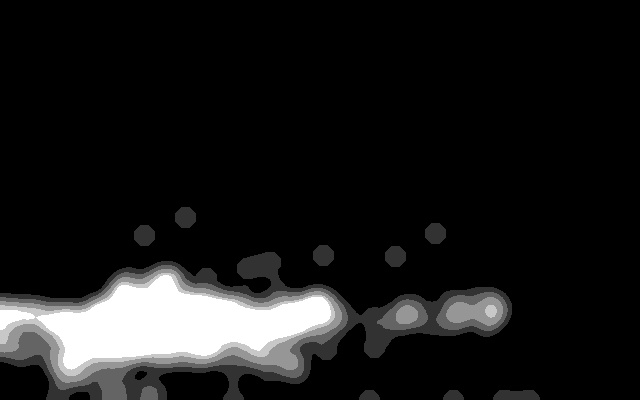}\label{figure::30N}}
     \caption{Heatmaps during 5, 15 and 30 seconds in both distracted (left) and neutral (right) driving scenarios.}
     \label{5_15_30}
\end{figure}

In accordance with the initial expectations, the variation of the obtained shapes is visually different between \textit{neutral} and \textit{distracted} driving (see Figure \ref{5_15_30}, columns a and b). These shapes occupy a wide area in neutral driving, as the driver checks his environment often. However, in the presence of cognitive distraction, the covered area narrows as the driver fixates more on a single zone.

For a heatmap gained by longer observation times, a better separable visual pattern is obtained. This is due to the fact that with a longer observation time, the driver has more time to explore his environment in neutral driving, whereas in distracted driving, as he often fixates on a narrowed zone, observing for a longer period does not greatly change the heatmap. As a result, the difference between neutral and distracted driving patterns becomes more obvious with a longer observation time (see window size in Figure \ref{5_15_30}). Nevertheless, safety-oriented solutions should warn of dangerous situations as quickly as possible.

\subsection{Scores} \label{subsubsection::Scores}

The relationships between the observed window size and the classification result are presented in Table \ref{window_size_accuracy}. A window of 5 seconds achieved 63\%  accuracy, whereas a window of 60 seconds achieved 85\% of accuracy. These results are in accordance with the expectations based on the previous heatmap observations (see Figure \ref{5_15_30}).

\begin{table}[htbp]
    \caption{Performances in terms of window size in seconds.}
    \begin{center}
    \begin{tabular}{|c|c|c|c}
    \hline
    \textbf{Window}&\multicolumn{2}{|c|}{\textbf{Performances}} \\
    \cline{2-3} 
    \textbf{Size} & \textbf{\textit{Accuracy}}& \textbf{\textit{F1-Score}} \\
    \hline
    5 & 62.582 & 0.609  \\
    10 & 68.082 & 0.669  \\
    15 & 70.778 & 0.698  \\    
    20 & 73.330 & 0.723  \\
    25 & 78.675 & 0.779  \\
    30 & 81.408 & 0.808  \\    
    35 & 83.286 & 0.827  \\
    40 & 84.655 & 0.876  \\
    45 & 84.805 & 0.875  \\
    50 & 85.814 & 0.849  \\
    55 & 85.166 & 0.810  \\
    60 & 85.286 & 0.827  \\  
    \hline
    \end{tabular}
    \label{window_size_accuracy}
    \end{center}
\end{table}

The presented results were obtained by averages of scores from 10 random training--testing splits (stratified k-fold cross validation) in which the subjects in the training sets were always distinct from the subjects in the testing set to prevent over-fitting. The confusion matrix obtained by averaging these folds, based on heatmaps of 30 seconds, is presented in  Table \ref{Table::30_sec_confusion_matrix}.

\begin{table}[htbp]
    \caption{Confusion matrix for the 30 second heatmap-based classifier (accuracy: 81.408\%, F1 Score: 0.808.)}
    \begin{center}
    \begin{tabular}{cc|cc}
        \multicolumn{1}{c}{} &\multicolumn{1}{c}{} &\multicolumn{2}{c}{Predicted} \\ 
        \multicolumn{1}{c}{} & 
        \multicolumn{1}{c|}{} & 
        \multicolumn{1}{c}{Neutral} & 
        \multicolumn{1}{c}{Distracted} \\ \hline
        \multirow{2}{*}{Actual}
        & Neutral  & 0.895 & 0.105   \\ 
        & Distracted  & 0.265   & 0.735 \\ \hline
    \end{tabular}
    \label{Table::30_sec_confusion_matrix}
    \end{center}
\end{table}

\section{Conclusion} \label{section::Conclusion}

The field of human-centered artificial intelligence is tackling its current issues and aims to increasingly assist humans in their daily life. Specifically, intelligent systems are now part of vehicles and assist the driver to increase road safety.

In this work, we have investigated the problem of the detection of the high cognitive load of drivers through an image-based representation created by tracking the driver's eye-gaze projection on an imaginary plane surface (heatmaps).

The variation of the obtained shapes revealed the driver's cognitive distraction. These shapes occupy a wide area in neutral driving, as the driver checks his environment often. However, in the presence of cognitive distraction, the covered area narrows, as the driver fixates more on a single zone (see Figure \ref{5_15_30}).

The trained \gls{svm}-based classifiers achieved 85.2\% accuracy; thus, the proposed method has good discriminative power between  neutral and distracted driving scenarios.

For a heatmap obtained by longer observation times, a better separable visual pattern was obtained. Nevertheless, safety-oriented solutions should warn of dangerous situations as quickly as possible. Thus, a window size compromise should be selected between the algorithmic performance and alerting time. For the real participants, we selected two classifiers working in parallel, with different window sizes. The first one classified with a short window size to warn of problems as fast as possible ($t=10 sec$), and the second one used a long window size in order not to miss any dangerous situations ($t=30 sec$).

\subsection{Further Discussions} \label{subsection::FurtherDiscussions}

Future work in this context should involve increasing the participant numbers and collecting more data; however, the scientific background, the obtained heatmap shapes for neutral and distracted driving and the used validation technique shows that this result could be generalized to a wider population.

Further studies should also include other road types and conditions, as in this work, the driver's cognitive load estimation was studied only under similar conditions (on highway roads with speed limited to 90 km/h, during the day-time, with low traffic and good weather conditions). Once more data are collected, further studies should investigate \gls{cnn}-based classifiers, and ablation tests per feature set should be presented.

Due to the end-user's needs, modern vehicles are equipped with only a single central \gls{nir} camera. In parallel with this demand, our method is based on a single central \gls{nir} camera. However, multiple cameras would open the possibility of implementing a wider and curved imaginary surface, which would increase the data availability.

The 3D View (see Figure \ref{figure::3DViewerGdi}) extracts other gaze-related features such as the mirror checking frequency. These data should also be added to the feature set.

\begin{figure}[!htb]
    \begin{center}
    \includegraphics[width=\linewidth]{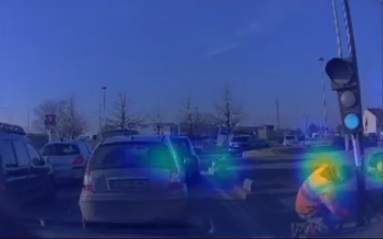} %[scale=.6]
    \end{center}
    \caption{Overlay of the generated heatmap and the front camera's view. We can observe the low activity for the front vehicle and the high activity for the worker next to the traffic light, the presence of which is an unexpected event.}
    \label{subplot::pedestrian_heatmap}
\end{figure}

Finally, the real positions of other vehicles, pedestrians and road signs could be taken into account in the heatmap creation process; additionally, we could change the weights for specific zones in the heatmap. Figure \ref{subplot::pedestrian_heatmap} shows an uncommon case, in which the expected heatmap would be different from the default ones.

\section*{ACKNOWLEDGMENT}

The authors would like to thank Kevin Nguyen for his help over the entire course of the project, to Omar Islas-Ramirez for reviewing this article, to all collaborators---Julien, Pantelis, Gaëlle, Vincenzo, Joao, Philippe, Emmanuel and Gabriel---and to the students from Sorbonne University---Antoine, Rodolphe, William and Anes.

%%%%%%%%%%%%%%%%%%%%%%%%%%%%%%%%%%%%%%%%%%%%%%%%%%%%%%%%%%%%%%%%%%%%%%%%%%%%%%%%

\Urlmuskip=0mu plus 1mu\relax
\bibliographystyle{./bibliography/IEEEtran}
\bibliography{./bibliography/bibliography}

\begin{thebibliography}{10}
\providecommand{\url}[1]{#1}
\csname url@samestyle\endcsname
\providecommand{\newblock}{\relax}
\providecommand{\bibinfo}[2]{#2}
\providecommand{\BIBentrySTDinterwordspacing}{\spaceskip=0pt\relax}
\providecommand{\BIBentryALTinterwordstretchfactor}{4}
\providecommand{\BIBentryALTinterwordspacing}{\spaceskip=\fontdimen2\font plus
\BIBentryALTinterwordstretchfactor\fontdimen3\font minus
  \fontdimen4\font\relax}
\providecommand{\BIBforeignlanguage}[2]{{%
\expandafter\ifx\csname l@#1\endcsname\relax
\typeout{** WARNING: IEEEtran.bst: No hyphenation pattern has been}%
\typeout{** loaded for the language `#1'. Using the pattern for}%
\typeout{** the default language instead.}%
\else
\language=\csname l@#1\endcsname
\fi
#2}}
\providecommand{\BIBdecl}{\relax}
\BIBdecl

\bibitem{who_report}
\BIBentryALTinterwordspacing
W.~H. Organization, ``Global status report on road safety 2018: Summary,''
  World Health Organization, Technical documents, 2018. [Online]. Available:
  \url{https://apps.who.int/iris/handle/10665/277370}
\BIBentrySTDinterwordspacing

\bibitem{nhtsa}
\BIBentryALTinterwordspacing
{United States Department of Transportation}, ``National highway traffic safety
  administration: Distracted driving,'' 2018, accessed: 2020-01-03. [Online].
  Available: \url{https://www.nhtsa.gov/risky-driving/distracted-driving}
\BIBentrySTDinterwordspacing

\bibitem{nhtsa_faq}
\BIBentryALTinterwordspacing
------, ``Policy statement and compiled faqs on distracted driving,'' accessed:
  2020-01-03. [Online]. Available:
  \url{http://www.nhtsa.gov.edgesuite-staging.net/Driving+Safety/Distracted+Driving/Policy+Statement+and+Compiled+FAQs+on+Distracted+Driving}
\BIBentrySTDinterwordspacing

\bibitem{US6496117B2}
\BIBentryALTinterwordspacing
``System for monitoring a driver's attention to driving,'' US Patent
  US6\,496\,117B2, 2001. [Online]. Available:
  \url{https://patents.google.com/patent/US6496117B2/en}
\BIBentrySTDinterwordspacing

\bibitem{SCHOOLER20141}
\BIBentryALTinterwordspacing
J.~W. Schooler, M.~D. Mrazek, M.~S. Franklin, B.~Baird, B.~W. Mooneyham,
  C.~Zedelius, and J.~M. Broadway, ``Chapter one - the middle way: Finding the
  balance between mindfulness and mind-wandering,'' in \emph{Psychology of
  Learning and Motivation}, B.~H. Ross, Ed.\hskip 1em plus 0.5em minus
  0.4em\relax Academic Press, 2014, vol.~60, pp. 1 -- 33. [Online]. Available:
  \url{http://www.sciencedirect.com/science/article/pii/B9780128000908000019}
\BIBentrySTDinterwordspacing

\bibitem{euroncap}
\BIBentryALTinterwordspacing
``{Euro NCAP 2025 Roadmap, In Pursuit of Vision Zero},'' Euro NCAP, Tech. Rep.,
  09 2017. [Online]. Available:
  \url{https://cdn.euroncap.com/media/30700/euroncap-roadmap-2025-v4.pdf}
\BIBentrySTDinterwordspacing

\bibitem{neisser2014cognitive}
\BIBentryALTinterwordspacing
U.~Neisser, \emph{Cognitive Psychology: Classic Edition}, ser. Psychology Press
  \& Routledge Classic Editions.\hskip 1em plus 0.5em minus 0.4em\relax Taylor
  \& Francis, 2014. [Online]. Available:
  \url{https://books.google.fr/books?id=WSGcBQAAQBAJ}
\BIBentrySTDinterwordspacing

\bibitem{atkinson_shiffrin_1977}
\BIBentryALTinterwordspacing
R.~Atkinson and R.~Shiffrin, ``Human memory: A proposed system and its control
  processes,'' in \emph{Human Memory}, G.~BOWER, Ed.\hskip 1em plus 0.5em minus
  0.4em\relax Academic Press, 1977, pp. 7 -- 113. [Online]. Available:
  \url{http://www.sciencedirect.com/science/article/pii/B9780121210502500065}
\BIBentrySTDinterwordspacing

\bibitem{Baddeley1974}
\BIBentryALTinterwordspacing
A.~D. Baddeley and G.~Hitch, ``Working memory,'' in \emph{Psychology of
  Learning and Motivation}, G.~H. Bower, Ed.\hskip 1em plus 0.5em minus
  0.4em\relax Academic Press, 1974, vol.~8, pp. 47 -- 89. [Online]. Available:
  \url{http://www.sciencedirect.com/science/article/pii/S0079742108604521}
\BIBentrySTDinterwordspacing

\bibitem{10.1007/s10648-005-3951-0}
\BIBentryALTinterwordspacing
J.~J.~G. Van~Merrienboer and J.~Sweller, ``Cognitive load theory and complex
  learning: Recent developments and future directions,'' \emph{Educational
  Psychology Review}, vol.~17, pp. 147--177, 06 2005. [Online]. Available:
  \url{https://doi.org/10.1007/s10648-005-3951-0}
\BIBentrySTDinterwordspacing

\bibitem{10.1007/978-3-030-14273-5_3}
\BIBentryALTinterwordspacing
G.~Orru and L.~Longo, \emph{The Evolution of Cognitive Load Theory and the
  Measurement of Its Intrinsic, Extraneous and Germane Loads: A Review}.\hskip
  1em plus 0.5em minus 0.4em\relax American Physiological Association, 02 2019,
  pp. 23--48. [Online]. Available:
  \url{https://psycnet.apa.org/doiLanding?doi=10.1037\%2F0022-0663.100.1.223}
\BIBentrySTDinterwordspacing

\bibitem{Nilsson2017}
\BIBentryALTinterwordspacing
E.~Nilsson, C.~Ahlstr\"om, S.~Barua, C.~Fors, P.~Lindén, B.~Svanberg,
  S.~Begum, M.~U. Ahmed, and A.~Anund, ``Vehicle driver monitoring –
  sleepiness and cognitive load,'' VTI, Tech. Rep., 2017, vTI rapport 937A.
  [Online]. Available:
  \url{http://vti.diva-portal.org/smash/get/diva2:1096341/FULLTEXT01.pdf}
\BIBentrySTDinterwordspacing

\bibitem{USOO6998972B2}
\BIBentryALTinterwordspacing
``Driving workload estimation,'' US Patent USOO6\,998\,972B2, 2002. [Online].
  Available: \url{https://patents.google.com/patent/US6998972B2/en}
\BIBentrySTDinterwordspacing

\bibitem{US6974414B2}
\BIBentryALTinterwordspacing
``System and method for monitoring and managing driver attention loads,'' US
  Patent US6\,974\,414B2, 2003. [Online]. Available:
  \url{https://patents.google.com/patent/US6974414B2/zh-TW}
\BIBentrySTDinterwordspacing

\bibitem{US10399575B2}
\BIBentryALTinterwordspacing
``Cognitive load driving assistant,'' US Patent US10\,399\,575B2, 2015.
  [Online]. Available: \url{https://patents.google.com/patent/US10399575B2/en}
\BIBentrySTDinterwordspacing

\bibitem{Sahadat2013}
\BIBentryALTinterwordspacing
M.~N. {Sahadat}, S.~{Consul-Pacareu}, and B.~I. {Morshed}, ``Wireless
  ambulatory {ECG} signal capture for {HRV} and cognitive load study using the
  neuromonitor platform,'' in \emph{2013 6th International IEEE/EMBS Conference
  on Neural Engineering (NER)}, 11 2013, pp. 497--500. [Online]. Available:
  \url{http://doi.org/10.1109/NER.2013.6695980}
\BIBentrySTDinterwordspacing

\bibitem{8417207}
\BIBentryALTinterwordspacing
M.~{Niemann}, A.~{Prange}, and D.~{Sonntag}, ``Towards a multimodal
  multisensory cognitive assessment framework,'' in \emph{2018 IEEE 31st
  International Symposium on Computer-Based Medical Systems (CBMS)}, June 2018,
  pp. 24--29. [Online]. Available:
  \url{https://doi.org/10.1109/CBMS.2018.00012}
\BIBentrySTDinterwordspacing

\bibitem{KUMAR201670}
\BIBentryALTinterwordspacing
N.~Kumar and J.~Kumar, ``Measurement of cognitive load in {HCI} systems using
  {EEG} power spectrum: An experimental study,'' \emph{Procedia Computer
  Science}, vol.~84, pp. 70 -- 78, 2016, proceeding of the Seventh
  International Conference on Intelligent Human Computer Interaction (IHCI
  2015). [Online]. Available:
  \url{http://www.sciencedirect.com/science/article/pii/S1877050916300825}
\BIBentrySTDinterwordspacing

\bibitem{7795957}
\BIBentryALTinterwordspacing
V.~{Alizadeh} and O.~{Dehzangi}, ``The impact of secondary tasks on drivers
  during naturalistic driving: Analysis of {EEG} dynamics,'' in \emph{2016 IEEE
  19th International Conference on Intelligent Transportation Systems (ITSC)},
  2016, pp. 2493--2499. [Online]. Available:
  \url{https://ieeexplore.ieee.org/document/7795957}
\BIBentrySTDinterwordspacing

\bibitem{vanderWel2018}
\BIBentryALTinterwordspacing
P.~van~der Wel and H.~van Steenbergen, ``Pupil dilation as an index of effort
  in cognitive control tasks: A review,'' \emph{Psychonomic Bulletin {\&}
  Review}, vol.~25, no.~6, pp. 2005--2015, Dec 2018. [Online]. Available:
  \url{https://doi.org/10.3758/s13423-018-1432-y}
\BIBentrySTDinterwordspacing

\bibitem{10.1145/1743666.1743701}
\BIBentryALTinterwordspacing
O.~Palinko, A.~L. Kun, A.~Shyrokov, and P.~Heeman, ``Estimating cognitive load
  using remote eye tracking in a driving simulator,'' in \emph{Proceedings of
  the 2010 Symposium on Eye-Tracking Research \& Applications}, ser. ETRA
  ’10.\hskip 1em plus 0.5em minus 0.4em\relax New York, NY, USA: Association
  for Computing Machinery, 2010, p. 141–144. [Online]. Available:
  \url{https://doi.org/10.1145/1743666.1743701}
\BIBentrySTDinterwordspacing

\bibitem{7535416}
\BIBentryALTinterwordspacing
Y.~{Liao}, S.~E. {Li}, G.~{Li}, W.~{Wang}, B.~{Cheng}, and F.~{Chen},
  ``Detection of driver cognitive distraction: An svm based real-time algorithm
  and its comparison study in typical driving scenarios,'' in \emph{2016 IEEE
  Intelligent Vehicles Symposium (IV)}, 2016, pp. 394--399. [Online].
  Available: \url{http://doi.org/10.1109/IVS.2016.7535416}
\BIBentrySTDinterwordspacing

\bibitem{10.1016/j.trf.2018.01.017}
\BIBentryALTinterwordspacing
P.~Biswas and G.~Prabhakar, ``Detecting drivers' cognitive load from saccadic
  intrusion.'' \emph{Transportation Research Part F: Traffic Psychology and
  Behaviour}, vol.~54, pp. 63--78, 2018. [Online]. Available:
  \url{https://doi.org/10.1016/j.trf.2018.01.017}
\BIBentrySTDinterwordspacing

\bibitem{Cora2016}
\BIBentryALTinterwordspacing
M.~P. Cora, ``Analyzing cognitive workload through eye related measurements: A
  meta-analysis,'' Master's thesis, Wright State University, Department of
  Biomedical, Industrial, and Human Factors Engineering, 2016. [Online].
  Available:
  \url{https://corescholar.libraries.wright.edu/etd_all/1507/?utm_source=corescholar.libraries.wright.edu\%2Fetd_all\%2F1507\&utm_medium=PDF\&utm_campaign=PDFCoverPages}
\BIBentrySTDinterwordspacing

\bibitem{WANG2014227}
\BIBentryALTinterwordspacing
Y.~Wang, B.~Reimer, J.~Dobres, and B.~Mehler, ``The sensitivity of different
  methodologies for characterizing drivers’ gaze concentration under
  increased cognitive demand,'' \emph{Transportation Research Part F: Traffic
  Psychology and Behaviour}, vol.~26, pp. 227 -- 237, 2014. [Online].
  Available:
  \url{http://www.sciencedirect.com/science/article/pii/S136984781400120X}
\BIBentrySTDinterwordspacing

\bibitem{Recarte2003MentalWW}
\BIBentryALTinterwordspacing
M.~A. Recarte and L.~M. Nunes, ``Mental workload while driving: effects on
  visual search, discrimination, and decision making.'' \emph{Journal of
  experimental psychology. Applied}, vol. 9 2, pp. 119--37, 2003. [Online].
  Available: \url{https://doi.org/10.1037/1076-898X.9.2.119}
\BIBentrySTDinterwordspacing

\bibitem{10.1145/3173574.3174226}
\BIBentryALTinterwordspacing
L.~Fridman, B.~Reimer, B.~Mehler, and W.~T. Freeman, ``Cognitive load
  estimation in the wild,'' in \emph{Proceedings of the 2018 CHI Conference on
  Human Factors in Computing Systems}, ser. CHI ’18.\hskip 1em plus 0.5em
  minus 0.4em\relax New York, NY, USA: Association for Computing Machinery,
  2018. [Online]. Available: \url{https://doi.org/10.1145/3173574.3174226}
\BIBentrySTDinterwordspacing

\bibitem{Purves2001}
\BIBentryALTinterwordspacing
D.~Purves, G.~Augustine, D.~Fitzpatrick, and al., \emph{Types of Eye Movements
  and Their Functions}, 2nd~ed.\hskip 1em plus 0.5em minus 0.4em\relax
  Sunderland: Neuroscience, 2001. [Online]. Available:
  \url{https://www.ncbi.nlm.nih.gov/books/NBK10991/}
\BIBentrySTDinterwordspacing

\bibitem{doi:10.1002/acp.3100}
\BIBentryALTinterwordspacing
B.~Park and R.~Brünken, ``The rhythm method: A new method for measuring
  cognitive load—an experimental dual-task study,'' \emph{Applied Cognitive
  Psychology}, vol.~29, no.~2, pp. 232--243, 2015. [Online]. Available:
  \url{https://onlinelibrary.wiley.com/doi/abs/10.1002/acp.3100}
\BIBentrySTDinterwordspacing

\bibitem{10.17077/drivingassessment.1563}
\BIBentryALTinterwordspacing
L.~Hsieh, S.~Seaman, and R.~Young, ``Eye glance analysis of the surrogate tests
  for driver distraction,'' in \emph{Proceedings of the Eighth International
  Driving Symposium on Human Factors in Driver Assessment, Training and Vehicle
  Design}, 06 2015, pp. 141--147. [Online]. Available:
  \url{https://ir.uiowa.edu/drivingassessment/2015/papers/22/}
\BIBentrySTDinterwordspacing

\bibitem{10.3389/fpsyg.2018.02208}
\BIBentryALTinterwordspacing
P.~D. Gajewski, E.~Hanisch, M.~Falkenstein, S.~Thönes, and E.~Wascher, ``What
  does the n-back task measure as we get older? relations between
  working-memory measures and other cognitive functions across the lifespan,''
  \emph{Frontiers in Psychology}, vol.~9, p. 2208, 2018. [Online]. Available:
  \url{https://www.frontiersin.org/article/10.3389/fpsyg.2018.02208}
\BIBentrySTDinterwordspacing

\bibitem{Valeo_dms}
Valeo, ``Driver monitoring: a camera to monitor driver alertness,''
  \url{https://www.valeo.com/en/driver-monitoring/}, march 2015.

\bibitem{scikit-learn}
\BIBentryALTinterwordspacing
F.~Pedregosa, G.~Varoquaux, A.~Gramfort, V.~Michel, B.~Thirion, O.~Grisel,
  M.~Blondel, P.~Prettenhofer, R.~Weiss, V.~Dubourg, J.~Vanderplas, A.~Passos,
  D.~Cournapeau, M.~Brucher, M.~Perrot, and E.~Duchesnay, ``Scikit-learn:
  Machine learning in {P}ython,'' \emph{Journal of Machine Learning Research},
  vol.~12, no.~85, pp. 2825--2830, 2011. [Online]. Available:
  \url{http://jmlr.org/papers/v12/pedregosa11a.html}
\BIBentrySTDinterwordspacing

\end{thebibliography}

\end{document}